# *Effect of Couette Flow on Electroconvective Vortices*

Yifei Guan and Igor Novosselov [§]

*Department of Mechanical Engineering, University of Washington, Seattle, U.S.A. 98195*

*Institute for Nano-Engineered Systems, University of Washington, Seattle, U.S.A. 98195*

**Abstract**—Numerical simulation of Electroconvective vortices behavior in the presence of Couette flow between two infinitely long electrodes is investigated. The two-relaxation-time Lattice Boltzmann Method with fast Poisson solver solves for the spatiotemporal distribution of flow field, electric field, and charge density. Couette cross-flow is applied to the solutions after the electroconvective vortices are established. Increasing cross-flow velocity deforms the vortices and eventually suppresses them when threshold values of shear stress are reached.

## I. INTRODUCTION

Insights into multiphysics interactions are crucial for understanding Electrohydrodynamics (EHD flows: (1) the electric field from the potential difference between the anode and cathode and its modifications by the space charge effects; (2) the ion motion in the electric field; (3) the interaction between the motion of ions and the neutral molecules; and (4) the inertial and viscous forces in the complex flow. As a subset of EHD, electroconvection (EC) is a phenomenon where convective transport is induced by unipolar discharge into a dielectric fluid [1-20]. The EC stability problem was first analyzed by a simplified non-linear hydraulic model [21, 22] and linear stability analysis without charge diffusion [23, 24]. Atten & Moreau [25] showed that in the weak-injection limit, $C<<1$, where $C$ is the charge injection level, the flow stability is determined by the criterion $T_c C^2$, where $T_c$ is the linear stability threshold for the electric Rayleigh number $T$ — a ratio between electric force to the viscous force. In the space-charge-limited (SCL) injection, $C \to \infty$, the flow stability is determined by $T_c$ only. The experimental observations [26, 27] have shown that, for the SCL scenario, $T_c = 100$, while linear stability analysis suggests $T_c = 160.45$ for the same conditions [25]. It was suggested that the discrepancy is due to the omission of the charge diffusion term in the analysis [28]. The effect of charge diffusion was investigated using linear stability analysis with a Poiseuille flow [11] and by non-linear analysis using a multiscale method [16]. It was shown that the charge diffusion has a non-negligible effect on $T_c$ and the transient behavior depends on the Reynolds number ($Re$) [11, 16].

To gain insight into the complexity of the EC flow, the problem can be investigated by numerical simulations. The earlier finite difference model simulations have shown that strong numerical diffusivity may contaminate the model [2]. Other numerical approaches include the particle-in-cell method [29], finite volume method with the flux-corrected transport scheme [30], total variation diminishing scheme [4, 7, 13-15], and the method of characteristics [3]. Lattice Boltzmann model (LBM) was shown to predict the linear and finite amplitude stability criteria of the subcritical bifurcation in the EC flow [17-20] for both 2D and 3D flow scenarios. This unified LBM transforms the elliptic Poisson equation into a parabolic reaction-diffusion equation and introduces artificial coefficients to control the evolution of the electric potential.

The EC stability problem is analogous to Rayleigh-Benard convection (RBC) [20, 31-36]. Of particular interest is the suppression of the RBC cells in the cross-flow [37]. A non-dimensional group $Gr/Re^2$, the ratio of buoyancy to the inertia force, was used to parametrize the effect of the applied shear, where $Gr$ is the Grashof number. For $Gr/Re^2 > 10$, the effect of the cross-flow is insignificant, while for $Gr/Re^2 < 0.1$, the effect of the buoyancy can be neglected. In the EC flow scenario, 2D finite volume simulations of Poiseuille flow show that the critical electric Rayleigh number, $T_c$, depends on the $Re$ and ion mobility parameter, $M$ [12].

In this paper, we investigate the EC stability in the cross-flow between two parallel electrodes. The segregated solver used in the study combined a two-relaxation-time LBM modeling fluid and charged species transport and a Fast Fourier Transform Poisson solver to solve for the electric field directly [38]. Couette cross-flow scenario provides shear stress. A subcritical bifurcation is described by the ratio of the electrical force to the viscous force.

[§] ivn@uw.edu

## II. METHODOLOGY

The governing equations for EHD flow include the Navier-Stokes equations (NSE) with the electric forcing term $\mathbf{F}_e = -\rho_c \nabla \varphi$ in the momentum equation, the charge transport equation, and the Poisson equation for electric potential.

$$\nabla \cdot \mathbf{u} = 0, \quad (1)$$

$$\rho \frac{D\mathbf{u}}{Dt} = -\nabla P + \mu \nabla^2 \mathbf{u} - \rho_c \nabla \varphi, \quad (2)$$

$$\frac{\partial \rho_c}{\partial t} + \nabla \cdot \left[ (\mathbf{u} - \mu_b \nabla \varphi) \rho_c - D_c \nabla \rho_c \right] = 0, \quad (3)$$

$$\nabla^2 \varphi = -\frac{\rho_c}{\varepsilon}, \quad (4)$$

where $\rho$ is the density, $\mu$ is the dynamic viscosity, $\mathbf{u} = (u_x, u_y)$ is the velocity vector field, $P$ is the static pressure, $\mu_b$ is the ion mobility, $D_c$ is the ion diffusivity, $\rho_c$ is the charge density, $\varepsilon$ is the electric permittivity, and $\varphi$ is the electric potential. The electric force provides a source term in the momentum equation (Eq. 2) [11, 39-41].

Non-dimensional analysis of the governing equations NSE (Eq. 1-4) yields:

$$\nabla^* \cdot \mathbf{u}^* = 0 \quad (5)$$

$$\frac{D^* \mathbf{u}^*}{D^* t^*} = -\nabla^* P^* + \frac{M^2}{T} \nabla^{*2} \mathbf{u}^* - CM^2 \rho_c^* \nabla^* \varphi^*, \quad (6)$$

$$\frac{\partial^* \rho_c^*}{\partial^* t^*} + \nabla^* \cdot \left[ (\mathbf{u}^* - \nabla^* \varphi^*) \rho_c^* - \frac{1}{Fe} \nabla^* \rho_c^* \right] = 0. \quad (7)$$

$$\nabla^{*2} \varphi^* = -C \rho_c^*, \quad (8)$$

where the asterisk denotes the non-dimensional variables. In the absence of cross-flow, non-dimensional governing equations yield four dimensionless parameters describing the system's state [4, 6, 7, 9, 11-20].

$$M = \frac{(\varepsilon/\rho)^{1/2}}{\mu_b}, \quad T = \frac{\varepsilon \Delta \varphi_0}{\mu \mu_b}, \quad C = \frac{\rho_0 H^2}{\varepsilon \Delta \varphi_0}, \quad Fe = \frac{\mu_b \Delta \varphi_0}{D_e}, \quad (9)$$

where $H$ is the distance between the electrodes (two infinite plates), $\rho_0$ is the injected charge density at the anode, and $\Delta \varphi_0$ is the voltage difference between the electrodes. The physical interpretations of these parameters are as follows: $M$ is the ratio between hydrodynamic mobility and the ionic mobility; $T$ is the ratio between electric force to the viscous force; $C$ is the charge injection level [11, 16]; and $Fe$ is the reciprocal of the charge diffusivity coefficient [11, 16].

## III. RESULTS

To model EC vortices, the hydrostatic base-state is perturbed using wave-form functions with a small amplitude that satisfies the boundary conditions and continuity equation:

$$u_x = L_x \sin(2\pi y / L_y) \sin(2\pi x / L_x) \times 10^{-3}$$
$$u_y = L_y \left[ \cos(2\pi y / L_y) - 1 \right] \cos(2\pi x / L_x) \times 10^{-3} \quad (10)$$

The physical domain size $Lx = 1.22m$ and $Ly = 1m$ limits the perturbation wavenumber to $\lambda_x = 2\pi / L_x \approx 5.15(1/m)$, yielding the most unstable mode under the conditions $C = 10, M = 10$ and $Fe = 4000$ [18]. The electric Nusselt number, $Ne = I / I_0$, serves as a flow stability criteria, where $I$ is the cathode current for a given solution and $I_0$ is the cathode current for the hydrostatic solution [4, 18]. For cases where EC vortices exist, $Ne > 1$. For a strong ion injection, the EC stability largely depends on $T$, so, in this analysis, $T$ is varied, while other non-dimensional parameters are held constant at $C = 10$, $M = 10$, and $Fe = 4000$ [42].

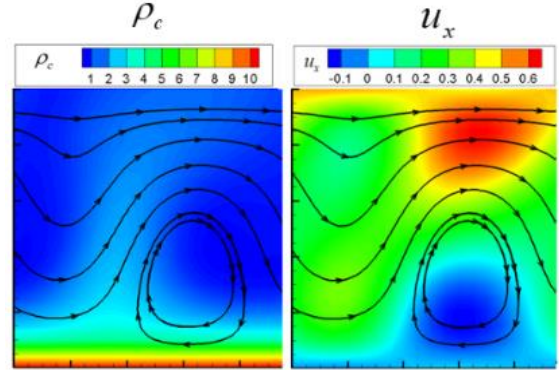

FIG. 1. Charge density and x-direction velocity contour of the EC with cross-flow. Couette flow with $u_{wall} = 0.5 m/s$; one of the two vortices is suppressed.

The Couette cross-flow is added after EC vortices are established by assigning constant velocity of the upper wall. FIG. 1 shows the charge density and x-direction velocity for Couette cross-flow ($u_{wall} = 0.5 m/s$). The Couette cross-flow stretches the vortices in the direction of the bulk flow and may eliminate one of the two vortices. The vortex suppression is due to the interaction of the vortex's x-velocity components with cross-flow; these interactions are the strongest near the walls where x-velocity are the greatest. For example, the clockwise vortex will be deformed at some oblique angle as in x-direction (streamwise) flow accelerates the upper region of the vortex and slows down the bottom region (relative to the mean velocity). This is reversed in the case of the counterclockwise rotating vortex. For strong cross-flow, both vortices are eliminated, and $I=I_0$, $Ne=1$. Here, the EC contribution to the flow field is negligible at higher values of shear stress (higher velocity), and the flow field is the same as the applied cross-flow.

FIG. 2 shows the extended stability analysis of EC with and without cross-flow [38] by introducing a finite velocity of the upper wall (cathode). For a constant $T$, $Ne$ decreases as $U_{wall}$ increases. The applied shear stress reduces the EC effect on the flow.

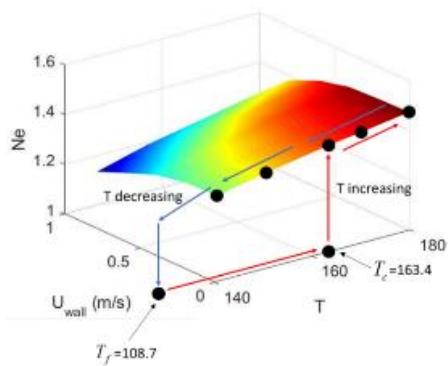

*FIG. 2. Electric Nusselt number depends on the electric Rayleigh number $T$ and applied velocity of the upper wall $U_{wall}$ for Couette type cross-flow.*

## IV. CONCLUSION

The 2D numerical study extends the EC stability analysis to Couette flow between two infinitely long parallel electrodes. The numerical approach utilizes the two-relaxation-time LBM to solve the flow and charge transport equations and a Fast Poisson Solver to solve the Poisson equation. Shear stress from the applied cross-flow deforms the EC vortices due to the interaction of streamwise velocity components resulting in vortex suppression at high crossflow velocities.

## ACKNOWLEDGMENT

This research was supported by the DHS Science and UK Home Office; grant no. HSHQDC-15-531 C-B0033, by the National Institutes of Health, grant NIBIB U01 EB021923 and NIBIB R42ES026532 subcontract to UW